\documentclass[onecolumn,usenatbib,usegraphicx]{mn2e}
\usepackage{amsmath}
\usepackage{amssymb}
\usepackage{lscape}
\def\be{\begin{equation}}
\def\ee{\end{equation}}

\def\mnras{MNRAS}
\def\apj{ApJ}

\def\apjs{ApJS}

\def\aap{A\&A}

\title[Phase-Averaged Gamma-Ray Spectra from MSPs]{Phase-Averaged Gamma-Ray Spectra from Rotation-Powered Millisecond Pulsars}
\author[Z. J. Jiang et al.]{Z. J. Jiang,$^{1,2}$ S. B. Chen,$^{1,3}$ X. Li,$^{1}$ and L. Zhang$^{1}$\thanks{E-mail: lizhang@ynu.edu.cn} \\
$^1$Department of Physics, Yunnan University, Kunming, 650091 China \\
$^2$Yunnan Observatories, Chinese Academy of Sciences, Kunming, 650011 China\\
$^3$Department of Physical Science and Technology, Kunming University, Kunming, 650214 China}

\begin{document}

%\title{Phase-Averaged Gamma-Ray Spectra from Rotation-Powered Millisecond Pulsars}
%
%\author{Z. J. Jiang$^{1,2}$, S. B. Chen$^{1}$, X. Li$^{1}$, and L. Zhang$^{1}$}
%%\affil{$^1$Department of Physics, Yunnan University, Kunming, China}
%\affil{$^1$Department of Physics, Yunnan University, Kunming 650091 China; \email{\myemail}}
%\affil{$^2$National Astronomical Observatories/Yunnan Observatory, Chinese Academy of Sciences, Kunming, China}
%%\affil{$^2$National Astronomical Observatories/Yunnan Observatory, Chinese Academy of Science, P.O. Box 110, Kunming 650011, China}
%%\affil{$^3$Kunming University, Kunming 650031, China}
\maketitle

\begin{abstract}
{\it Fermi}-LAT has detected pulsed gamma-ray emissions with high
confidences from more than 40 millisecond pulsars (MSPs). Here we
study the phase-averaged gamma-ray properties of MSPs by using
revised version of a self-consistent outer gap model. In this
model, a strong multipole magnetic field near the stellar surface
for a MSP is assumed and such a field will be close to the surface
magnetic fields ($\sim 10^{11}- 10^{12}$ G) of young pulsars; the
outer gap of a MSP is controlled by photon-photon pair production
process, where the effects of magnetic inclination angle
($\alpha$) and magnetic geometry have been taken into account,
therefore the fractional size of the outer gap is a function of
not only pulsar's period and magnetic field strength but also
magnetic inclination angle and radial distance  to the neutron
star, the inner boundary of the outer gap can be estimated by the
pair production process of the gamma-ray photons which are
produced by the back-flowing particles through the null charge
surface; inside the outer gap, a Gaussian distribution of the
parallel electric field along the trans-field thickness is
assumed, and the gamma-ray emission is represented by the emission
from the average radial distance along the central field lines of
the outer gap. Using this model, the phase-averaged gamma-ray
spectra are calculated and compared with the observed spectra of
37 MSPs given by the second {\it Fermi}-LAT catalog of gamma-ray
pulsars, our results show that the {\it Fermi}-LAT results can be
well explained by this model. The thermal X-ray emission properties
from MSPs are also investigated.

\end{abstract}
\section{Introduction}
It is believed that millisecond pulsars (MSPs) are recycled
neutron stars which spin up by accreting mass from their
companions. MSPs are characterized by shorter spin periods ($P<30$
ms) and smaller spin-down rates ($\dot{P}<10^{-17} \rm s ~s^{-1}$)
compared to the non-recycled young pulsars, indicating lower
values of surface dipole magnetic field strengthes and lower
spin-down powers. Although these objects had been studied
extensively at radio wavelength, it was not clear whether MSPs
could produce gamma-rays like those of young pulsars before the
launch of {\it Fermi Gamma-ray Space Telescope}.

The second {\it Fermi} Large Area Telescope (LAT) catalog of 117
$\gamma$-ray pulsars with high-confidences has been published
recently \citep{abdo13}, which consists of 77 (42 radio-loud and
35 radio-quiet) young or middle-aged $\gamma$-ray pulsars and 40
millisecond $\gamma$-ray pulsars.  Therefore, MSPs have became an
important class of GeV objects, but the origin of gamma-ray
photons from MSPs is still debated. From the {\it Fermi}-LAT
observations of the MSPs \citep[e.g.,][]{abdo13}, the observed
spectra can be fitted by a power law with an exponential cut-off,
where the spectral index and the cut-off energy changes with the
MSPs. Compared to young pulsars, it has been found that the MSPs
share the same properties with young pulsars, so the gamma-ray
emission mechanism for MSPs is similar with those of young
pulsars.

For either a young pulsar or a MSP, it is generally believed that
the plasma with a charge density $\rho_{\rm GJ}\approx
-\vec{\Omega}\cdot \vec{B}/2\pi c $ \citep{GJ69} will surround a
rotating neutron star (NS) with an angular velocity $\Omega$ and
co-rotate with the NS within the light cylinder with a radius
$R_{\rm L}=c/\Omega$. {The global flow of plasma in the
pulsar magnetosphere will result in charge depletion regions
(called as gaps) in the open field line regions}. Inside a gap, a strong electric field
which is parallel to the magnetic field (called as an accelerating
electric field or a parallel electric field) can be produced, the
accelerating electric field will accelerate charged particles to
relativistic energies and then produce gamma-ray emission through
synchrotron radiation, curvature radiation and inverse Compton
scattering from lower-energy photons. Due to different
accelerating regions, various models have been proposed to explain
the pulsed gamma-ray emission from young pulsars, for examples,
polar cap model \citep[e.g.,][]{rs75,Daugherty82}, slot gap model,
\citep[e.g.,][]{arons83,muslimov03,muslimov04,Harding08}, annular
gap model \citep[e.g.,][]{qiao04,Du09}, and outer gap model
\citep[e.g.,][]{CHRa,CHRb,romani96,zc97,zh04,ZL09,H13,Li13}. For
the high energy emissions from the MSPs, different models have
also been proposed, such as annular gap model \citep{Du10,Du13},
outer gap model \citep{tak12}, two layer outer gap model
\citep{wang10}, and the emission geometry of MSPs has been studied
by using the light curve from {\it Fermi}-LAT
\citep{Venter09,venter12}.  It should be noted that current
observations by the {\it Fermi}-LAT tend to favor the models in
which gamma rays are generated in the outer magnetosphere
\citep{abdo10a,abdo10b}. Therefore we will focus on the outer gap
model for high energy emissions from the MSPs in this paper.

Compared to young pulsars, MSPs have shorter spin periods and
smaller surface dipole magnetic fields. As suggested by
\citet{zc03}, if there is only a dipole magnetic field in the
magnetosphere of the MSP, then it is difficult to create the outer
gap to accelerate $e^\pm$ pairs to relativistic energy in the
outer gaps. Therefore,  \citet{zc03} assumed that a strong
multipole magnetic field exits near the stellar surface of the NS
although a global dipole magnetic field can describe the magnetic
field far from the star well. In this case, thermal X-rays are
produced by the outer gap heating NS surface, and will collide
with high energy gamma-rays produced in the outer gap through
curvature radiation inside the outer gap, producing enough $e^\pm$
pairs to sustain steady outer gaps. In the self-sustained outer
gap model of \citet{zc03}, however, they did not consider the
effects of the magnetic inclination angle and magnetic geometry
and assumed that the typical radiation region locates at half of
the light cylinder ($R_{\rm L}/2$). \citet{zh07} revised the model
of \citet{zc03} by taking the effects of the inclination angle and
the magnetic geometry into account, and found that the fractional
size of outer gap is a function not only of the pulsar period and
magnetic field  but also of the radial distance to the NS and the
inclination angle. \citet{zh07} also assumed that the average
properties of high-energy photon emission from the outer gap can
be approximated by the radiation at an average radius.

Recently, \citet{Li13} described a revised version of the outer
gap for explaining $\gamma$-ray properties of young $\gamma$-ray
pulsars. In this outer gap model, an outer gap exists in the open
field line region, where the lower boundary along the trans-field
direction is in the last open field lines and the upper boundary
is constrained by the photon-photon pair production process or
critical field lines which are perpendicular to the rotational
axis at the light cylinder, resulting in two possible types of
outer gaps for the young pulsars, and the vertical distribution of
the parallel electric field can be approximated as a Gaussian
distribution (see details in \citet{Li13} or references therein).
Using this model, \citet{Li13} reproduced the observed
phase-averaged spectra of young gamma-ray pulsars. In this paper,
we construct a revised model of the outer gap for studying the
phase-averaged properties of gamma-ray emission from MSPs. The
paper is organized as follows. In Section 2, we describe the
details of the revised model. In Section 3, we apply the outer gap
model to fit the phase-averaged gamma-ray emission and study the
gamma-ray properties of MSPs. In Section 4, we give a brief
conclusions and discussions.

\section{Model Description}

As mentioned above, the observations indicate that high-energy
properties of the MSPs are similar to those of young pulsars, so
we believe that some features of the outer gap model for young
pulsars given by \citet{Li13} can be also used to describe high
energy emissions from the MSPs. In this paper, we will assume that
the shape of the outer gap for a young pulsar described by
\citet{Li13} is also available for a MSP. Below we describe the
details of the outer gap model used in this paper.

\subsection{Accelerating Electric Field and Thermal X-Rays from Outer Gap Heating}

The issue of the accelerating electric field inside the outer gap
is a complicated issue. In the classic outer gap model (or vacuum
outer gap model) \citep{CHRb}, the outer gap is assumed to be thin
and to extend from the null charge surface to the light cylinder.
In such a gap, the accelerating electric field has a distribution
along the direction of the gap height (or trans-field direction).
If  $h(r)$ and $z$ represent the local thickness of the outer gap
and a variable along the trans-field direction, respectively,
\citet{CHRb} demonstrated that the accelerating electric field is
proportional to $q(1-q)$ and reaches a maximum at $q=1/2$, where
$q=z/h(r)$. On the other hand, the variations of the accelerating
electric field along the direction of the magnetic fields for
young pulsars have been investigated by self-consistently solving
the Poisson equation for electrical potential, and the Boltzmann
equations of electrons/positrons, and gamma-rays in both vacuum
and nonvacuum outer gaps \citep[e.g.,][]{HS99,Tet06,LZ09}.

According to \citet{CHRb}, \citet{zc97} used the maximum
accelerating electric field in the vacuum dipole magnetic field as
the approximation of the accelerating electric field at position
($r$, $\theta$), where $r$ is the radial distance to the star and
$\theta$ is the polar angle. This accelerating electric field at
position ($r$, $\theta$) which is above the null charge surface
can be approximated as \citep[e.g.,][]{zc97,zh04}
\begin{equation}
E_{\parallel\rm max} (r,\theta)=f_{\rm
m}^2(r)B(r,\theta)\left(\frac{s}{R_{\rm L}}\right)\;\; {\rm for}\;
 r_{\rm n}\le r \le r_{\rm out}\;, \label{Epara}
\end{equation}
where $f_m(r)\equiv h(r)/R_{\rm L}$ is the fractional size of
outer gap, $s$ is the local curvature radius, $R_{\rm L}$ is the
radius of the light cylinder, and $B(r,\theta)$ is the dipole
magnetic field strength at the position ($r$, $\theta$). In Eq.
(\ref{Epara}), $r_{\rm n}$ is the radial distance of the null
charge surface and corresponding polar angle is
$\theta_n=\tan^{-1}[(3\tan\alpha +\sqrt{8+9\tan^2\alpha})/2]$.
Below the null charge surface, the accelerating electric field
below the null charge surface can be approximated by
\citep[e.g.,][]{Tet08,ZL09}
\begin{equation}
E^{\rm in}_{\parallel \rm max}=E_{\parallel\rm max}(r_{\rm
n},\theta_{\rm n})\frac{(r/r_{\rm in})^2-1}{(r_{\rm n}/r_{\rm
in})^2-1}\;\rm{for}~~ r_{\rm in}\le r<r_{\rm n}\;,
\end{equation}
where $r_{\rm in}$ is the radial distance of the inner boundary
(see Eq. (9) of \citet{Li13}).

Since the maximum accelerating electric field locates at the
center of the gap height for a given position ($r$, $\theta$), so
we can choose the magnetic field lines which cross the center of
the gap height as typical field lines of accelerating particle and
then emitting gamma-rays, i.e., the accelerating electric field at
any position ($r$, $\theta$) of the typical field line being
tangent to the light cylinder with a radius of $R_{\rm L}'=\kappa
R_{\rm L}$ ($\kappa\ge 1$) is given by Eq. (\ref{Epara}). In this
case, we can use a Gaussian distribution to approximate the distribution of the
accelerating electric field along the trans-field diraction given
by \citet{Li13}, which is
\begin{eqnarray}
\nonumber
  E_{\parallel p}(q)&=&\frac{E_{\parallel \rm max}-E_{\parallel \rm
  min}}{1-e^{-\frac{1}{2}0.5^2/\sigma^2_g}}\times\\
  &&\left[
e^{-\frac{(q-0.5)^2}{2\sigma_g^2}}+\frac{E_{\parallel \rm
min}-e^{-\frac{1}{2}0.5^2/\sigma^2_g}E_{\parallel \rm
max})}{E_{\parallel \rm max}-E_{\parallel \rm
  min}}\right], \label{Ep}
\end{eqnarray}
where $q=z/h(r)$, $E_{\parallel \rm  min}=(2/3)(e/R_{\rm
L}^2)(s/R_{\rm L})^{-2}$ is the minimum value of the accelerating
field line along the trans-field direction, and $\sigma_g$ is
{the standard deviation of the Gaussian distribution} which is
treated as a model parameter \citep{Li13}.

We now consider the thermal X-rays from the outer gap heating. In
the self-sustained outer gap model \citep{zc03,zh07}, the local
magnetic field will dominate over the global dipole field in the
region $R$ to $R+\delta r$, but the magnetic field can be
described by the dipole field in the region $r>R+\delta r$, where
$\delta r$ is the distance above the stellar surface in which the
local magnetic field is equal to the dipole magnetic field and $R$
is NS radius. Assuming a localized dipole field form of the local
magnetic field, the relationship between the local field and the
dipole field can be expressed as $B^0_s/B_d^0=[(R+\delta
r)/R]^{-3}[(l+\delta r)/l]^{3}$ \citep[e.g.,][]{zc03}, where
$B^0_d=3.2\times 10^{19}(P\dot{P})^{0.5}$ G is the dipole magnetic
field at the stellar surface, and $l$ is the typical radius of
curvature of the local magnetic field with a surface strength
$B^0_s$.
%In such a case, \citet{zh07} investigated the thermal
%X-ray production due to the outer gap heating of the stellar
%surface and obtained the average X-ray energy of thermal X-rays
%from outer gap heating, i.e., $<E_X>\approx 2.7 k T^{(1)}_{\rm
%m}$, where $k$ is the Boltzmann constant and $T^{(1)}_m$ is given
%by equation (11) of \citet{zh07}.
{
In such a case, Zhang \& Cheng (2003) and Zhang et al. (2007) investigated the
thermal X-ray production caused by the back-flowing particles
from the outer gap (also see Takata et al. 2012). The return particle flux can be
approximated by $\dot{N}_{e^\pm}\simeq 0.5f(r_{\rm in})\dot{N}_{\rm GJ}$, where $f(r_{\rm in})$
is the fractional size of the outer gap in the inner boundary and $\dot{N}_{\rm GJ}$ is the
Goldreich-Julian current. Those particles loss energies
via curvature radiation and the remaining energy can be approximated as
$E(R+\delta r)=m_ec^2\gamma(R+\delta r)$ when reaching the distance $R+\delta r$,
and $E(R)=m_ec^2\gamma(R)$ when reaching the NS surface.
In the region between $R<r<R+\delta r$, the local magnetic field dominate the global
dipole field, and the curvature photons will convert into $e^\pm$ pairs
in the strong local magnetic field, and finally deposit on the NS surface with an effective
area $A_{\rm eff}^{(1)}\simeq \pi (\delta r)^2$ and produce the first component of thermal X-ray
emission, and the luminosity can be approximated as $L_{X}^{\rm th1}\simeq 0.5m_ec^2 [\gamma(R+\delta r)-\gamma(R)]\dot{N}_{e^\pm}$,
and the temperature can be expressed as $T_m^{(1)}=(L_{X}^{\rm th1}/A_{\rm eff}^{(1)}\sigma_{SB})^{1/4}$,
which is given by equation (11) of \citet{zh07}.
The second component of thermal X-rays is produced by the back-flowing particle with energy
$E(R)=m_ec^2\gamma(R)$ deposited on an effective area $A_{\rm eff}^{(2)}\simeq (B_d^0/B_s^0)\pi r_{\rm pc}^2$
with thermal luminosity $L_{X}^{\rm th1}\simeq m_ec^2 \gamma(R)\dot{N}_{e^\pm}$ and temperature
$T_m^{(2)}$ which is given by equation (13) of \citet{zh07}.
}

\subsection{The Fractional Size of the Outer Gap}
The thermal X-ray photons with energy $<E_X>\approx 2.7 k T^{(1)}_{m}$ will collide with the curvature gamma-ray photons
with energy $E_\gamma$ given by equation (7) of \citet{zh07} to produce enough $e^\pm$ to sustain a steady outer gap,
where $k$ is the Boltzmann constant.
With photon-photon pair production condition, \citet{zh07} derived
the expression of the fractional size of the outer gap above the
null charge surface, which is
\begin{equation}
f_{\rm m}(r,\alpha)\approx 5.9\times 10^{-2} P^{26/21}_{\rm ms}
\left({B^0_{\rm d}\over 10^8~{\hbox{G}}}\right)^{-4/7}\delta
r^{2/7}_5\eta(\alpha, r)\;\;{\rm for}\; r_{\rm n}\le r\le r_{\rm
out}\;, \label{fsize}
\end{equation}
where $P_{\rm ms}$ is the pulsar period in units of millisecond, $B^0_{\rm
d}$ is dipole magnetic field at stellar surface in units of Gauss,
$\delta r_5=\delta r / 10^5$ cm, and the function $\eta (r,
\alpha)$ contains the effects of inclination angle and magnetic
geometry, and can be estimated by using Eqs. (17) - (20) of
\citet{zh07}.
{
The distance $\delta r$, where the local magnetic field
is equal to the dipole magnetic field, can be estimated as
$\delta r=l[(B_s^0/B_d^0)^{1/3}-1]/[1-(l/R)(B_s^0/B_d^0)^{1/3}]$.
We can see that $\delta r$ depends on $l$ and $B_s^0/B_d^0$, both of
them are still unclear. Generally, it is assumed that the curvature
radius $l$ of the local magnetic field is of the order of the crust thickness of
the star, $l\sim 10^5 $cm, and $B_s^0\sim 10-10^3 B_d^0$ \citep[e.g.,][]{rud91,chen98}.
In our following calculation, we assume that $l=0.5\times10^5$ cm, $B_s^0/B_d^0=300$ and $R$=12 km
as taken by \citet{zh07}, then we have $\delta r_5\sim 4$.
}

In the region from the inner boundary to the null charge surface,
the fractional size of the outer gap can be estimated by using the
magnetic flux conservation in the outer gap \citep[e.g.,][]{ZL09},
i.e., $f^{\rm in}_{\rm m}(r,\alpha)=f_{\rm m}(r_{\rm
n},\alpha)[B(r_{\rm n}, \theta_{\rm n})/B(r, \theta)]^{1/2}$ for
$r_{\rm in}\le r\le r_{\rm n}$.

\subsection{Phase-Averaged Gamma-Ray Spectrum}

In this section, we consider the phase-averaged spectrum of
gamma-ray emission from a MSP. \citet{zc03} (also see
\citet{zh07}) did not considered the accelerating electric field
along the trans-field direction and assumed the accelerating
electric field along a specific magnetic field line near the last
open field line as typical one for simplicity. Different from
\citet{zh07}, in this paper, the typical field line is the field
line which is tangent to the light cylinder with a radius of
$R_L'=\kappa R_{\rm L}$. Moreover we assume that high-energy
emission at an average radius $r_{\rm a}$ along the typical field
line can represents the typical emission of high-energy photons
from a MSP and the average radius over the entire outer gap is
given by
\begin{equation}
r_{\rm a}={\int^{r_{\rm out}}_{r_{\rm in}}f_{\rm m}(r,\alpha)rdr
\over \int^{r_{\rm out}}_{r_{\rm in}}f_{\rm m}(r,\alpha)dr}\;,
\label{rave}
\end{equation}
where $r_{\rm out}=\min (r_{\rm b},r_{\rm LC})$.

We assume that gamma-rays from a MSP are mainly produced by the
curvature radiation of the accelerated particles and the gamma-ray
beam points toward the observer. Therefore, the phase-averaged
gamma-ray spectrum of a young pulsar given by \citet{Li13} can
also used to describe that of a MSP. In this case, the
phase-averaged gamma-ray spectrum can be estimated as
\citep{Li13}:
\begin{equation}
F^{\rm th}_{\gamma}(E_{\gamma})\approx
\left(\frac{\Delta\phi}{\Delta\Omega}\right) \frac{f^3_{\rm
g}(r_a,\alpha)L_{\rm sd}}{E_\gamma d^2}U(x)\left(\frac{r_a}{R_{\rm
L}}\right)^{-4}\;, \label{Ftheo}
\end{equation}
where $d$ is the distance to the pulsar, $(r_{\rm a},\theta_a)$
represents the typical radiation position, $\Delta \phi$ is the
extension angle at azimuthal direction, $\Delta \Omega$ is the
gamma-ray beaming solid angle, $L_{\rm sd}\approx
3.8\times10^{35}(P/\rm ms)^{-4}(B^0_{\rm d}/{10^8 ~\rm G})^2$
erg/s is the spin down power of a pulsar, and the function $U(x)$
is given by Eq. (22) of \citet{Li13}.

For comparison with the observed cutoff energy of a MSP's SED, we
give the characteristic energy of the curvature photons, which is
\begin{eqnarray}
E_{\gamma}(r_{\rm a})&\approx & 4.3\times 10^{10}f^{3/2}_{m}(r_{\rm a})\left({B^0_{\rm d}\over
10^8{\rm {G}}}\right)^{3/4}P^{-7/4}_{\rm ms}\nonumber\\
&&\left(1-{3\over4}{a(\alpha)r_{\rm a}\over R_{\rm L}
}\right)^{3/8} \left({r_{\rm a}\over R_{\rm
L}}\right)^{-13/8}W^{-1/4}(r_{\rm a},\alpha) \;{\rm {eV}}\;,
\label{Ecura}
\end{eqnarray}
where $P_{\rm ms}= P/1$ ms, and $W(r_a,\alpha)=s/\sqrt{r_a R_{\rm
L}}=(4/3)[1-(3/4)a(\alpha)(r_a/R_{\rm
L}]^{3/2}/[\sqrt{a(\alpha)}(1-(1/2)a(\alpha)r_a/R_{\rm L})]$ with
$a(\alpha)=\sin \theta_{\rm LC}\sin^2(\theta_{\rm LC}-\alpha)$.

\section{Applications}

We now apply the above outer gap model to study the phase-averaged
gamma-ray spectra from MSPs. In the second {\it Fermi}-LAT catalog of
gamma-ray pulsars, there are 40 gamma-ray MSPs and the spectral
fitting results by using an exponentially cut-off power-law are
presented, and the observed spectral energy distributions (SEDs)
of these MSPs except for three of them (PSRs J1024-0719,
J1741+1351 and J1939+2134) are also presented. Here we compare the
model results with observed SEDs of 37 MSPs. The periods, surface
magnetic field and distances of these MSPs are listed in Table 1,
which will be used in our calculations.

For a pulsar with known period and period derivative, if the
inclination angle $\alpha$ is given, we can estimate the
fractional size of the outer gap, and then calculate the gamma-ray
SED. In fact, observationally it is difficult to obtain the value
of inclination angle, so we treat $\alpha$ as a model parameter
which can be determined by fitting the observed gamma-ray
spectrum. Two other parameters in our calculations are the
extension at azimuthal direction $\Delta \phi$ and beaming solid
angle $\Delta \Omega$. $\Delta \phi$ can be estimated by the pair
production, and $\Delta \Omega= 4\pi f_\Omega$, where $f_\Omega$
is the gamma-ray beaming fraction, which depends on the
inclination angle, viewing angle and the detail structure of the
outer gap \citep{watters09}, and the pulsar beaming model
indicated that $f_\Omega\sim 1$. In this paper, we do not consider
the specific value of $f_\Omega$ for each pulsar, we combine
$\Delta \phi$ and $\Delta\Omega$ as a single model parameter
$\frac{\Delta \phi}{\Delta \Omega}$ because $F^{\rm
th}_\gamma(E_\gamma)\propto (\Delta \phi/\Delta \Omega)$ (see Eq.
({\ref{Ftheo})). Therefore, for a given MSP, i.e., its rotation
period, period derivative and distance to the Earth are given, the
model has three parameters: $\alpha$, $\Delta \phi/\Delta \Omega$,
and $\sigma_{\rm g}$.

We use the outer gap model to reproduce the phase-averaged
gamma-ray spectra of MSPs observed by {\it Fermi}-LAT. In order to
determine the best model parameters, the chi-square goodness of
fit test is used, the chi-square test statistic can be written as
\begin{equation}
\chi^2=\sum_i \frac{[F_\gamma^{\rm ob}(E_\gamma)-F_\gamma^{\rm th}(E_\gamma)]^2}{\sigma^2(E_\gamma)}~,
\label{chi2}
\end{equation}
where $F_\gamma^{\rm th}(E_\gamma)$ is the model flux given by
equation ({\ref{Ftheo}), $F_\gamma^{\rm ob}(E_\gamma)$ is the
observed flux, and $\sigma(E_\gamma)$ is the corresponding error,
the observed data are taken from the second {\it Fermi}-LAT catalog of
gamma-ray pulsars \citep{abdo13}.

We fit the observed spectra as follows. First, we use a specific
value of $\alpha$ to estimate the fractional size $f_{\rm g}(r,
\alpha)$, then estimate the averaged radius $r_{\rm a}$ by
equation (\ref{rave}), and the corresponding fractional size
$f(r_{\rm a},\alpha)$. Second, we obtain an electronic field
distribution by equation (\ref{Ep}) with a specific value of
$\sigma_{\rm g}$. As proposed by \citet{Li13}, $\sigma_{\rm g}$
increases with the fractional size $f_{\rm g}(r_{\rm a},\alpha)$.
It is reasonable to assume that $\sigma_{\rm g}$ is larger for a
pulsar with thick outer gap than one with thin outer gap. In our
following calculation, for simplicity, we assume $\sigma_{\rm
g}=0.15$ when $f_{\rm g}(r_{\rm a},\alpha)<0.2$, $\sigma_{\rm
g}=0.2$ when $0.2\le f_{\rm g}(r_{\rm a},\alpha)<0.3$,
$\sigma_{\rm g}=0.25$ when $0.3\le f(r_a,\alpha)<0.4$, and
$\sigma_{\rm g}=0.3$ when $f_{\rm g}(r_{\rm a},\alpha)>0.4$. As
shown in our following calculations, the above assumption about
$\sigma_{\rm g}$ is a good approximation. Third, we calculate the
gamma-ray spectrum at typical radius $r_{\rm a}$ by equation
(\ref{Ftheo}) and compare it with the observation to find the best
fitting results of $\Delta \phi /\Delta \Omega$. It should be
noted that, the slope of the calculated spectrum is mainly
controlled by $\alpha$ and $\sigma_{\rm g}$, and the magnitude can
be modified by $\Delta \phi /\Delta \Omega$. So for a given
$\alpha$ and $\sigma_{\rm g}$, we change the value $\Delta \phi
/\Delta \Omega$ to get a minimum value of $\chi^2$ by equation
(\ref{chi2}). Finally, we change the values of $\alpha$ to repeat
above processes to obtained the minimum of $\chi^2$, then we can
obtain the best fitting parameters of $\alpha$ and $\Delta \phi
/\Delta \Omega$, and the reduced chi-square, $\chi^2/\nu$, where
$\nu=N-2$ is the degree of freedom and $N$ is the number of
observed data.

The best fitting parameters are presented in Table
\ref{Fermi_msp_tab1}, which include the inclination angle $\alpha$
and $\Delta \phi / \Delta \Omega$. We can see that the inclination
angles for $\sim 80\%$ MSPs are larger the $60^\circ$. In our
model, larger value of the inclination angle for a given MSP means
larger value of the $\gamma$-ray typical energy. For the best fits
of the parameter $\Delta \phi / \Delta \Omega$, change range is
very large, from $\sim 0.02$ to $\sim 2.8$, 7 of 33 MSPs have
$\Delta \phi / \Delta \Omega> 1$ which seems not to be reasonable.
By definition, $\Delta \phi$ is the extension angle of the outer
gap at the azimuthal direction and could be less than $2\pi$ and
 $\Delta \phi \sim 250^{\circ}$ for most
of young pulsars \citep{Li13}, $\Delta \Omega$ is the solid angle
of $\gamma$-ray beam, according to \citet{watters09},  the
gamma-ray beaming fraction $f_\Omega \sim 1$ for most young
$\gamma$-ray pulsars, which means $\Delta \Omega \sim 4\pi$.
Therefore, $\Delta \phi / \Delta \Omega \sim 0.35$ for a MSP with
$\Delta\phi\sim 250^\circ$ and $\Delta \Omega \sim 4\pi$. However,
as shown in Table \ref{Fermi_msp_tab1}, there are large
differences among the fitting value of $\Delta \phi / \Delta
\Omega$, which may be caused by the the uncertainties of distances
and the geometry of outer gaps. In fact, the inclination angle and
$\Delta \Omega$ for a given MSP can be estimated by reproducing
its high-quality $\gamma$-ray light curve in the frame of some
three-dimensional (3D) model (for example outer gap (OG) model, or
two-pole caustic (TPC) model), \citet{watters09} calculated the
beaming patterns and light curves of young $\gamma$-ray pulsars in
both TPC and OG models and found that $\gamma$-ray light curve
shapes depend sensitively on $\alpha$ and viewing angle $\zeta$.
\citet{Venter09} computed the $\gamma$-ray beaming patterns and
light curves of MSPs in both TPC and OG models and compared
calculation results with those observed by {\it Fermi}-LAT, they
gave the values of $\alpha$ and $f_{\Omega}$ for 7 MSPs (see Table
3 of \citet{Venter09}). Compared to our results, the differences
are obvious.

In Table \ref{Fermi_msp_tab1}, we list the values of the averaged
radius $r_{\rm a}$ and the coefficient $\kappa $ for each MSP. We
can see that the values of $r_{\rm a}$ for most MSPs are close to
1, indicating typical emissions occur near outer magnetosphere. On
the other hand, the values of $\kappa$ changes from $\sim 1.2$ to
$\sim 2.0$, showing that the curvature radiation regions for most
MSPs are close to the null charge surface. In this table, the
fractional size $f_{\rm m}$ estimated by equation (\ref{fsize})
are also listed.

The comparisons of model SEDs with the best fitting parameters
with observed ones for 37 MSPs are shown in Fig. \ref{fig1} and
Fig. \ref{fig2}. Although the observed SEDs of 37 MSPs are
available, the observed data are poor for some MSPs. Using a power
law with an exponential cutoff, i.e. $F^{\rm ob}_\gamma
(E_\gamma)\propto E^{-\Gamma}\exp(-E/E_{\rm cut})$, where $\Gamma$
is the spectral index and $E_{\rm cut}$ is the cutoff energy,
\citet{abdo13} have given the best-fit values of $\Gamma$, $E_{\rm
cut}$ and $F^{\rm ob}_\gamma (E_\gamma)$. In order to compare our
results with the observed data better, we also show the best fits
(blue lines) with one $\sigma$ uncertainties (red dashed lines) by
using a power law with an exponential cutoff given by
\citet{abdo13}. In Fig. \ref{fig1} and Fig. \ref{fig2}, our best
fitting model spectra with the fitting parameters listed in Table
1 are shown with the black solid lines, the data points are the
observed data from the {\it Fermi}-LAT, which are taken from
\citet{abdo13}. We can see that the outer gap model can well
explain the observations of most of MSPs, and the fitted
$\chi^2/\nu$ also indicate that the model results are consistent
with the observations. As shown in Fig. \ref{fig1} and Fig.
\ref{fig2}, however, 8 of 37 MSPs have large values of
$\chi^2/\nu> 10$, we believe that these large values are mainly
caused by the limited observation data. For example, there are
only 6 data points (including 3 upper limits ) for J1658-5324,
resulting in $\chi^2/\nu\approx 63$, same case is for J1648-5324;
for J2047+1053, 7 data points include 6 upper limits, resulting in
$\chi^2/\nu\approx 53$. Although large $\chi^2/\nu$ for these
MSPs, our model results can compare with the best fits by using a
power law with an exponential cutoff. On the other hand, from Fig.
\ref{fig1}, Fig. \ref{fig2}, and Table \ref{Fermi_msp_tab1}, 5
MSPs (J1514-4946, J1614-2230, J2043+1711, J2124-3358, and
J2302+4442) with better data points have $\chi^2/\nu$ values from
$\sim 5$ to $\sim 9$. The differences between our model results
with the data points can be divided two cases: the first case is
that our results cannot fit the data points at low energy part
(J1514-4946 and J2043+1711), implying that some radiation
mechanism may have a role in the low energy part; the second case
is that our results cannot fit the data points at peak energy
range of $E^2 F$ (J1614-2230, J2124-3358, and J2302+4442). In
fact, if we relax the limit of $\sigma_{\rm g}$ (see above), our
results can fit the data points well for the second case.

It should be noted the gamma-ray spectral properties of PSR
J0218+4232. This pulsar has high X-ray luminosity, hard power-law
shaped X-ray spectra and emits narrow X-ray pulse,
\citet{kuiper00} reported the likely detection of pulsed gamma-ray
emission from this pulsar using CGRO EGRET data and found that the
X-ray and gamma-ray pulse profile shapes (doubled peaks) are
similar. Since PSR J0218+4232 and Crab pulsar have the
similarities of high energy pulse profiles and have that the
magnetic field strengths near the light cylinders are comparable,
\citet{kuiper00} pointed out that pulsed high-energy nonthermal
emission from PSR J0218+4232 and Crab pulsar have a similar origin
in the pulsar magnetosphere. However, the gamma-ray pulse profile
observed by {\it Fermi}-LAT show a single pulse \citet{abdo13},
which is not consistent with that given by \citet{kuiper00}. For
the Crab pulsar, the synchrotron-self Compton radiation process
play an important role in the outer gap model
\citep[e.g.,][]{zc02,lz10}. Here, we find that only curvature
radiation cannot explain the observed SED of PSR J0218+4232 very
well (see Fig. \ref{fig1} and Table \ref{Fermi_msp_tab1}) and
other radiation mechanisms seem to be required.

{
In the outer gap model, there are two components of thermal X-ray emission
caused by heating of return particles from the outer gap. The observed
thermal X-ray spectra of MSPs can be fitted by blackbody radiation with one or two
temperatures. In Table \ref{Fermi_msp_tab1}, we list the observed
temperatures $T_{m,o}^{(1)}$ and $T_{m,o}^{(2)}$ for each MSP if available
from literatures, we also present the predicted temperatures $T_{m}^{(1)}$
and $T_{m}^{(2)}$ estimated by equations (11) and (13) of \citet{zh07},
respectively. We can see that $T_{m}^{(1)}\sim (0.5-0.8)\times 10^6$ K and
$T_{m}^{(2)}\sim (1.5-4)\times 10^6$ K, which are roughly consistent with
the observations. We also listed the observed thermal X-ray luminosities in
the seventh column, which are estimated by $L_X^{(\rm ob)}=4\pi d^2F_{\rm BB}$,
where $F_{\rm BB}$ is the unabsorbed X-ray flux, and the
modeled thermal luminosities $L_X^{(\rm th)}$ are listed in
seventeenth column, which are estimated by equation (27) of \citep{zh07}.
We can see that the model results can roughly explain the observations, and
the differences for some MSPs are likely caused by the limited observations
or by the uncertainties of multipole field.}

Finally, we compare model results of typical energy given by Eq.
(\ref{Ecura}) with the values of $E_{\rm cut}$ given by
\citet{abdo13}. In this revised outer gap model, we have assumed
that curvature photon emission at the position ($r_{\rm a}$,
$\theta_{\rm a}$) of the typical field line represents the typical
emission of high energy $\gamma$-rays from a given MSP, so the
characteristic energy estimated by  Eq. (\ref{Ecura}) for each MSP
would be a good approximation of observed $E_{\rm cut}$ because of
the feature of the curvature radiation. In Fig. \ref{fig3}, we
show the comparison of the model results with the observed $E_{\rm
cut}$ given by \citet{abdo13}, where dashed line represents
$E_{\rm cur}=E_{\rm cut}$ and solid line is the fit result. We can
see that the typical gamma-ray energy $E_{\rm cur}$ is comparable
with $E_{\rm cut}$, which satisfies the relation $E_{\rm
cut}=(0.37\pm 0.27)+(0.77\pm0.12)E_{\rm cur}$.

\section{Conclusions and Discussion}

We have studied the phase-averaged gamma-ray spectra from
rotation-powered MSPs based on the revised outer gap model of
\citet{zh07}. In this model, there is a strong multipole magnetic
field near the stellar surface, and the back-flowing particles
from the outer gap will finally convert into X-ray photons in this
strong magnetic field, which can collide with the curvature
gamma-ray photons in the outer gap and then produce
electon/positron pairs to sustain a self-consistent outer gap, and
the fractional size of the outer gap can be estimated by the pair
production condition. In this revised outer gap model, the effects
of the magnetic inclination angle and magnetic geometry have been
taken into account, and the gamma-ray emission at an averaged
radius can be used to represent the phase averaged gamma-ray
emission. In this paper, we have made three assumptions for study
of the gamma-ray emission from MSPs: (i) the gamma-ray emission
along the central field lines in the outer gap can be used to
represent the typical gamma-ray emission from MSPs, which is more
reasonable than the assumption that the gamma-ray emission along
the last open field lines represents the typical gamma-ray
emission; (ii) the typical gamma-ray emission at the averaged
radius $r_a$ can be used to represent the whole gamma-ray emission
from MSPs; (iii) the electric field in the trans-field direction
of the outer gap can be represented by a Gaussian distribution,
such an assumption has been successfully used to explain the
gamma-ray emission from normal young pulsars \citep{Li13}. With
this model, we calculated the phase-averaged gamma-ray spectra of
37 MSPs given by the second {\it Fermi}-LAT catalogue of gamma-ray
pulsars. The Chi-square goodness of fit test has been used to find
the best fitting parameters for each MSP. The estimated
phase-averaged gamma-ray spectra of these MSPs are presented in
Figures \ref{fig1} and \ref{fig2}, and the best fitting parameters
are listed in Table 1. We can see that the revised outer gap model
can well explain the observed phase-averaged gamma-ray spectra and
thermal X-ray emission of most MSPs.

In this paper, we also estimated the inner boundary of outer gap.
In the traditional outer gap model of gamma-ray pulsars, the inner
boundary of the outer gap is assumed at the null-charge surface.
However, \citet{H08} pointed out that the inner boundary of the
outer gap can be shifted inward from the null charge surface. Such
a modification is important when the light curve and
phase-resolved spectra of the normal pulsars is reproduced in the
three dimensional outer gap models \citep[e.g.][]{Tet08}. In this
work, we estimated the inner boundary of the outer gap of MSPs by
the pair production process and found that such an improvement
just causes marginal changes when the phase-averaged gamma-ray
spectra is calculated. However, it could play an important role
when the three dimensional out gap model of MSPs is constructed.

In our model, two parameters of $\alpha$ and
$\Delta\phi/\Delta\Omega$ can be estimated by fitting the
phase-averaged spectra of gamma-ray emissions from MSPs. These two
parameters can also estimated through simulating the light curve
of a MSP observed by {\it Fermi}-LAT. Generally, the inclination
angle is hard to measure directly and the simulation of the
$\gamma$-ray light curve for a pulsar needs to be performed in a
specific three-dimensional model. Simulated light curve depends on
not only the inclination angle $\alpha$ and viewing angle $\zeta$
but also the models used. For example, \citet{watters09} simulated
the $\gamma$-ray light curves for young pulsars in TPC and OG
models, the values of $\alpha$ and $\zeta$ for a young pulsar can
be different in different models (see their Table 1). Moreover,
the flux correction factor $f_{\Omega}$ and then $\Delta\Omega$
are also different for various model. The same case occurs in the
simulations of the $\gamma$-ray light curves of MSPs
\citep{Venter09,venter12}. Therefore, the values of $\alpha$ and
$\Delta\Omega$ obtained from the simulations of the $\gamma$-ray
light curves for $\gamma$-ray pulsars are still uncertain.
{Not only the magnetic inclination angle $\alpha$, but also the viewing angle
$\zeta$ are highly unknown. Thus, the flux correction factor (or equivalently
$\Delta\phi/\Delta\Omega$ in this paper) is anyway highly unknown as well.
There are 7 MSPs (out of total 37) with $\Delta\phi/\Delta\Omega>1$
as listed in Table \ref{Fermi_msp_tab1}. This is probably because the
flux correction factor becomes less than unity; that is, we are observing
such MSPs from the direction toward which more gamma-rays are emitted from other directions.
}

In our model, the role of the parameter
$\sigma_{\rm g}$ is to adjust the width of the accelerating
electric field along the trans-field direction and then has a
effect on the SED's shape: smaller value of $\sigma_{\rm g}$
results in a rise of the SED at low energy part for a MSP with a
given $\alpha$, vice versa. Our model indicate that the change of
$\sigma_{\rm g}$ is from $\sim 0.15$ to $\sim 0.30$.

{It should be noted that the properties of the multipole magnetic field play
important roles when estimate the fractional size $f_m$ and the thermal X-ray
temperatures $T_m^{(1)}$ and $T_m^{(2)}$, which are severely depend on the
unclear parameters $l$ and $B_s^0/B_d^0$. For simplicity, we take $l=0.5\times10^5$
cm and $B_s^0=300B_d^0$ for each MSP, which result in $\delta r_5\simeq 3.95$.
As shown by equation (\ref{fsize}), $f_m \propto \delta r_5^{2/7}$, then the gamma-ray luminosity
$L_\gamma \propto \delta r_5^{6/7}$. So different values of $l$ and $B_s^0/B_d^0$ will affect
the estimation about the gamma-ray luminosity, also influence the X-ray emission properties.
For example, if $l=1\times 10^5$ cm and $B_s^0/B_d^0=100$ \citep{zc03}, then $\delta r_5 \simeq 6.8$,
which can cause the difference when estimated the gamma-ray luminosity. However, such uncertainty
can be combined with uncertainty of the gamma-ray flux correction factor ($\delta \phi /\Delta \Omega$)
mentioned above and unchange the main results in this paper.}

{\it Fermi}-LAT second source catalog lists 575 unassociated
sources which have no obvious counterparts at other wavelengths,
some statistical methods have been used to analyze the origin of
these unassociated sources \citep[e.g.,][]{ackermann12}. Generally
speaking, most of the unassociated Galactic objects are likely
related to pulsars, pulsar wind nebulae, and supernova remnants.
Since these objects have pulsar-like spectra and variability
characteristics, the deep radio observation have been performed
and recognized more than 40 gamma-ray MSPs
\citep[e.g.,][]{Ray12,kerr12,keith11,ran11,bar13}, These
discoveries greatly increase the known population of MSPs, and the
deep survey by radio telescope and multi-wavelength observations
will find more MSPs, which will help us to improve our model.

\section*{Acknowledgments}
{We thank the anonymous referee for his/her valuable comments.
This work is partially supported by the National
Natural Science Foundation of China (NSFC 11173020, 11363006) and Doctoral
Fund of Ministry of Education of China (RFDP 20115301110005).}

\begin{figure}
\begin{center}
\includegraphics[width=150mm]{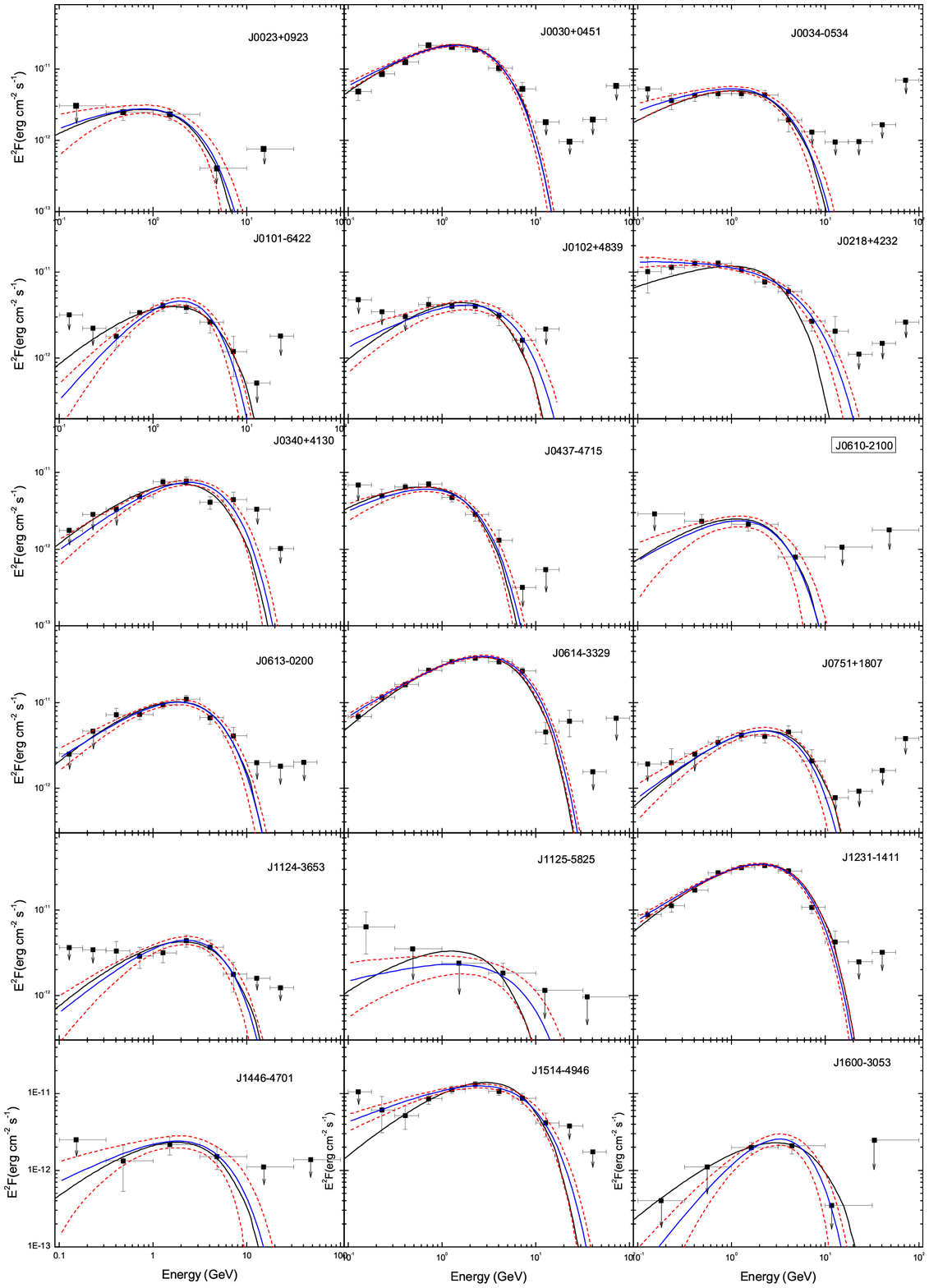}
%\vspace{3.5cm}
\caption{
The model results of
gamma-ray MSPs. The black solid lines represent the best fitting
model spectra with the fitting parameters listed in Table 1. The
data points are the observed data from the {\it Fermi}-LAT, which are
taken from \citet{abdo13}. The best fits (blue lines) with one
$\sigma$ uncertainties (red dashed lines) are also shown by using
a power law with an exponential cutoff \citep{abdo13} for
comparison.
}
\label{fig1}
\end{center}
\end{figure}

\clearpage
\begin{figure}
\begin{center}
%\vspace{3.5cm}
\includegraphics[width=150mm]{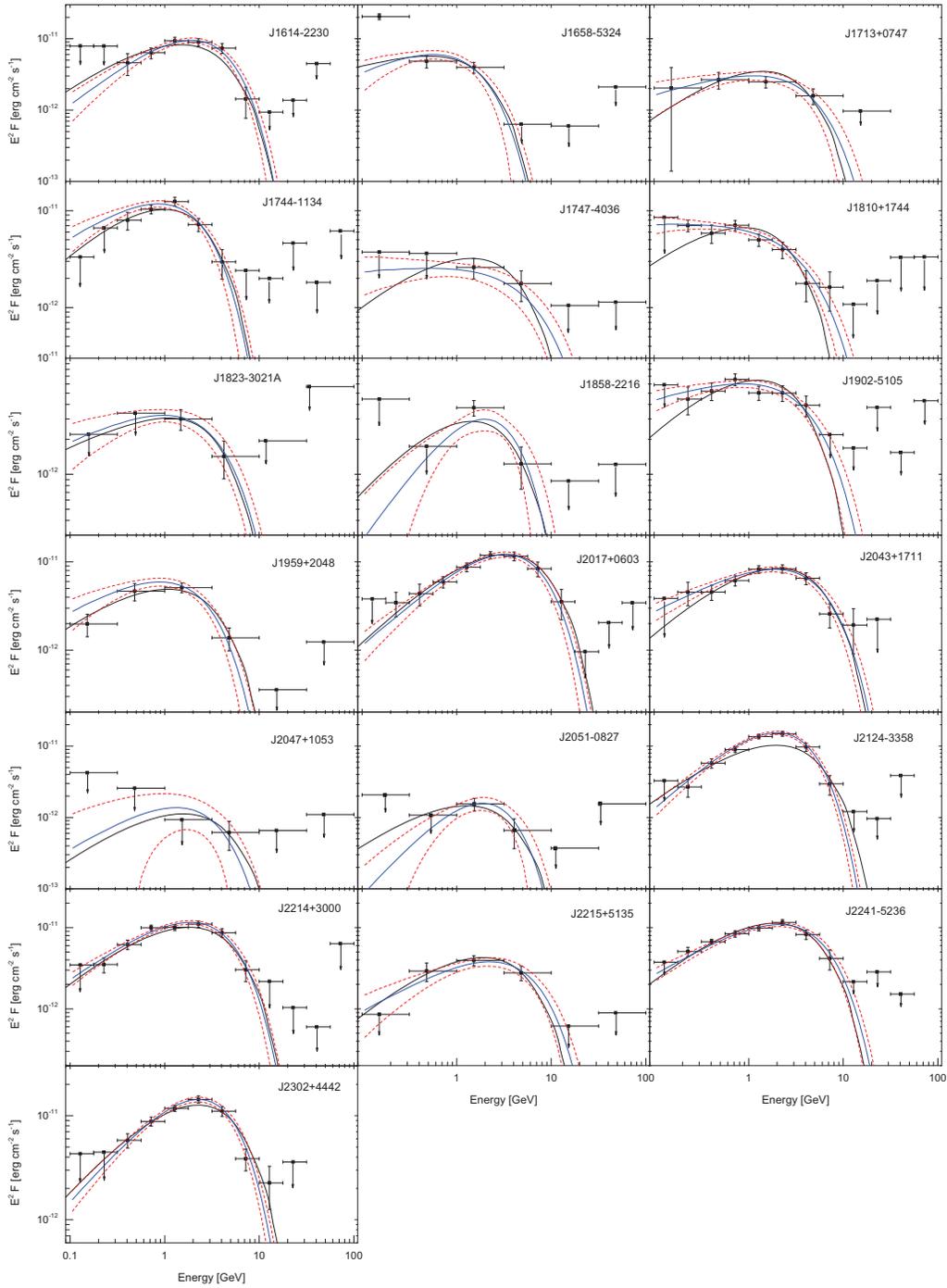}
\caption{Same as Figure \ref{fig1}.}
\label{fig2}
\end{center}
\end{figure}

\clearpage
\begin{figure}
\begin{center}
%\vspace{3.5cm}
\includegraphics[width=150mm]{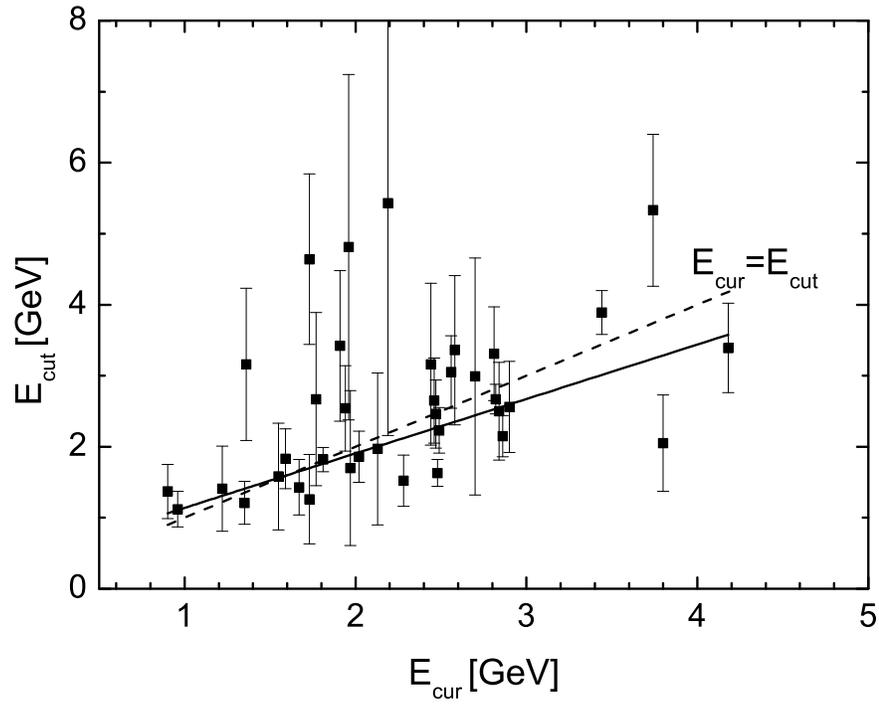}
\caption{Relation between the observed cutoff energy $E_{\rm cut}$ and the typical gamma-ray energy $E_{\rm cur}$. The solid
line indicate the fitting relation, the dashed line represents $E_{\rm cur}=E_{\rm cut}$.}
\label{fig3}
\end{center}
\end{figure}

\clearpage

\begin{landscape}
\begin{table}
\begin{center}
\caption{Observed quantities and model parameters of MSPs}\label{Fermi_msp_tab1}
\begin{tabular}{ccccccccccrcccccc}
\hline\hline

\multicolumn{1}{c}{} &
\multicolumn{6}{c}{Observed Quantities} &
\multicolumn{8}{c}{Model Parameters} &
\\
%\multicolumn{4}{c}{Deduced Parameters}\\

Name        & P     &$\log(B)$   & Dist & $T_{m,o}^{(1)}$ & $T_{m,o}^{(2)} $ &$\log(L_{X}^{\rm ob})$        &$\alpha$  &$f_{\rm m}$&$\kappa$ &$r_a$&$\sigma_g$&$\Delta \phi / \Delta \Omega$ &$\chi/\nu$ & $T_{m}^{(1)}$ & $T_{m}^{(2)}$ & $\log(L_{X}^{\rm th})$   \\
(PSR)        & (ms)  & (\rm G)      &(kpc)    &($10^6 \rm K)$     &($10^6 \rm K)$ & ($\rm erg~ s^{-1}$)  &(Deg) &           &       & ($R_L$)&         &                              &           &($10^6 \rm K$)  & ($10^6 \rm K$)   &($\rm erg ~s^{-1}$)   \\
%\tableline
\hline
J0023+0923    &  3.05  & 8.27  &  $0.69_{-0.11}^{+0.21}$ &   --     & ${2.2}^{(a)}$   & 30.0 & 48 & 0.27 &    1.36 &   0.80 & 0.20  &    0.11  & 0.12  & 0.78 & 2.89&  31.00 \\
J0030+0451    &  4.87  & 8.35  &  $0.28_{-0.06}^{+0.10}$ &  1.18    &$2.51^{(b)}$     & 30.6 & 58 & 0.48 &    1.75 &   0.85 & 0.30  &    0.11  & 2.81  & 0.71 & 2.49&  30.70 \\
J0034$-$0534  &  1.88  & 7.99  &  $0.54_{-0.10}^{+0.10}$ &   --     & $ 2.2^{(c)}$    & 29.6 & 65 & 0.27 &    1.32 &   0.94 & 0.20  &    0.09  & 0.30  & 0.69 & 2.34&  30.74 \\
J0101$-$6422  &  2.57  & 8.05  &  $0.55_{-0.08}^{+0.09}$ &   --     &       --        & --   & 69 & 0.39 &    1.56 &   0.95 & 0.25  &    0.05  & 0.57  & 0.65 & 2.12&  30.51 \\
J0102+4839    &  2.96  & 8.27  &  $2.32_{-0.43}^{+0.50}$ &   --     &       --        & --   & 73 & 0.36 &    1.53 &   0.97 & 0.25  &    0.81  & 0.21  & 0.64 & 2.17&  30.48 \\
J0218+4232    &  2.32  & 8.63  &  $2.64_{-0.64}^{+1.08}$ &   --     &   $2.9^{(d)}$   & 31.2 & 76 & 0.18 &    1.20 &   1.02 & 0.15  &    2.77  & 2.70  & 0.69 & 2.67&  30.76 \\
J0340+4130    &  3.30  & 8.15  &  $1.73_{-0.30}^{+0.30}$ &   --     &       --        & --   & 67 & 0.45 &    1.72 &   0.93 & 0.30  &    0.81  & 3.00  & 0.71 & 1.97&  30.64 \\
J0437$-$4715  &  5.76  & 8.76  &$0.156_{-0.001}^{+0.001}$& 0.52     &   $1.4^{(c)}$   & 30.5 & 32 & 0.28 &    1.37 &   0.75 & 0.20  &    0.02  & 0.90  & 0.86 & 3.75&  31.22 \\
J0610$-$2100  &  3.86  & 8.34  &  $3.54_{-1.00}^{+5.46}$ &  --      &    --           & --   & 53 & 0.34 &    1.51 &   0.83 & 0.25  &    1.88  & 0.82  & 0.75 & 2.75&  30.88 \\
J0613$-$0200  &  3.06  & 8.24  &  $0.90_{-0.20}^{+0.40}$ &  --      &    --           & --   & 72 & 0.39 &    1.58 &   0.97 & 0.25  &    0.27  & 1.02  & 0.64 & 2.15&  30.48 \\
J0614$-$3329  &  3.15  & 8.38  &  $1.90_{-0.35}^{+0.44}$ & --       & $2.67^{(e)}$    & 31.6 & 81 & 0.41 &    1.63 &   1.05 & 0.25  &    2.12  & 1.89  & 0.56 & 1.87&  30.15 \\
J0751+1807    &  3.48  & 8.22  &  $0.40_{-0.10}^{+0.20}$ & --       & ${1.69}^{(f)}$  & 29.8 & 72 & 0.47 &    1.78 &   0.97 & 0.30  &    0.02  & 0.31  & 0.63 & 2.06&  30.41 \\
J1124$-$3653  &  2.41  & 8.08  &  $1.72_{-0.36}^{+0.43}$ & --       & $ 5.2^{(a)}$    & 31.0 & 76 & 0.39 &    1.58 &   1.00 & 0.25  &    0.34  & 0.65  & 0.59 & 1.92&  30.31 \\
J1125$-$5825  &  3.10  & 8.64  &  $2.62_{-0.37}^{+0.37}$ & --       & --              & --   & 75 & 0.25 &    1.31 &   1.00 & 0.20  &    0.62  &39.68  & 0.68 & 2.55&  30.66 \\
J1231$-$1411  &  3.68  & 8.45  & $0.438_{-0.05}^{+0.05}$ & --       & $2.44^{(e)}$    & 30.5 & 76 & 0.40 &    1.62 &   1.00 & 0.25  &    0.15  & 2.47  & 0.62 & 2.14&  30.40 \\
J1446$-$4701  &  2.19  & 8.17  &  $1.46_{-0.22}^{+0.22}$ & --       & --              & --   & 79 & 0.34 &    1.43 &   1.06 & 0.25  &    0.11  & 0.03  & 0.59 & 1.93&  30.29 \\
J1514$-$4946  &  3.59  & 8.42  &  $0.94_{-0.12}^{+0.12}$ & --       & --              & --   & 81 & 0.46 &    1.76 &   1.05 & 0.30  &    0.19  & 5.29  & 0.56 & 1.85&  30.13 \\
J1600$-$3053  &  3.60  & 8.27  &  $1.63_{-0.27}^{+0.31}$ & --       & --              & --   & 78 & 0.52 &    1.93 &   1.02 & 0.30  &    0.12  & 0.09  & 0.58 & 1.85&  30.19 \\
J1614$-$2230  &  3.15  & 8.25  &  $0.65_{-0.05}^{+0.05}$ & --       & ${1.5}^{(g)}$   & 30.4 & 66 & 0.36 &    1.52 &   0.92 & 0.25  &    0.15  & 5.55  & 0.73 & 2.15&  30.73 \\
J1658$-$5324  &  2.44  & 8.22  &  $0.93_{-0.13}^{+0.11}$ & --       & --              & --   & 38 & 0.20 &    1.24 &   0.77 & 0.15  &    0.71  &63.36  & 0.83 & 3.19&  31.22 \\
J1713+0747    &  4.57  & 8.30  &  $1.05_{-0.05}^{+0.06}$ & --       & --              & --   & 57 & 0.47 &    1.72 &   0.85 & 0.30  &    0.25  &19.26  & 0.71 & 2.48&  30.71 \\
J1744$-$1134  &  4.07  & 8.29  &$0.417_{-0.017}^{+0.017}$& --       & ${3.31}^{(f)}$  & 29.0 & 46 & 0.37 &    1.53 &   0.79 & 0.25  &    0.16  & 1.47  & 0.76 & 2.81&  30.91 \\
J1747$-$4036  &  1.65  & 8.18  &  $3.39_{-0.76}^{+0.76}$ & --       & --              & --   & 79 & 0.24 &    1.26 &   1.08 & 0.20  &    1.04  &16.08  & 0.61 & 2.07&  30.44 \\
J1810+1744    &  1.66  & 7.95  &  $2.00_{-0.28}^{+0.31}$ &--        &$ 4.3^{(a)}$     &31.2  & 62 & 0.24 &    1.26 &   0.93 & 0.20  &    1.80  &14.18  & 0.72 & 2.45&  30.85 \\
J1823$-$3021A &  5.44  & 9.64  &  $7.60_{-0.40}^{+0.40}$ &--        & --              & --   & 82 & 0.16 &    1.18 &   1.06 & 0.15  &    2.48  & 0.42  & 0.71 & 3.25&  30.76 \\
J1858$-$2216  &  2.38  & 7.99  &  $0.94_{-0.13}^{+0.20}$ &--        & --              & --   & 65 & 0.36 &    1.50 &   0.93 & 0.25  &    0.13  & 9.54  & 0.67 & 2.20&  30.61 \\
J1902$-$5105  &  1.74  & 8.10  &  $1.18_{-0.21}^{+0.21}$ &--        & --              & --   & 74 & 0.25 &    1.28 &   1.02 & 0.20  &    0.36  & 3.78  & 0.65 & 2.20&  30.59 \\
J1959+2048    &  1.61  & 8.22  &  $2.49_{-0.49}^{+0.16}$ &--        & ${3.11}^{(f)}$  & 30.9 & 74 & 0.19 &    1.20 &   1.03 & 0.20  &    1.21  & 3.56  & 0.67 & 2.40&  30.72 \\
J2017+0603    &  2.90  & 8.20  &  $1.57_{-0.15}^{+0.15}$ & --       & --              & --   & 82 & 0.49 &    1.82 &   1.06 & 0.30  &    0.47  & 1.28  & 0.53 & 1.66&  29.98 \\
J2043+1711    &  2.38  & 8.07  &  $1.76_{-0.32}^{+0.15}$ & --       & --              & --   & 76 & 0.39 &    1.57 &   1.00 & 0.25  &    0.72  & 9.47  & 0.59 & 1.92&  30.32 \\
J2047+1053    &  4.29  & 8.48  &  $2.05_{-0.29}^{+0.32}$ & --       & --              & --   & 66 & 0.39 &    1.60 &   0.91 & 0.25  &    0.17  &53.22  & 0.72 & 2.42&  30.71 \\
J2051$-$0827  &  4.51  & 8.39  &  $1.04_{-0.15}^{+0.15}$ & --       & $2.90^{(h)}$    & 29.8 & 53 & 0.39 &    1.63 &   0.82 & 0.25  &    0.09  & 1.71  & 0.74 & 2.73&  30.84 \\
J2124$-$3358  &  4.93  & 8.51  &  $0.30_{-0.05}^{+0.07}$ & 0.5      &${2.2}^{(c)}$    & 30.3 & 67 & 0.45 &    1.77 &   0.91 & 0.30  &    0.03  & 7.27  & 0.70 & 2.36&  30.64 \\
J2214+3000    &  3.12  & 8.34  &  $1.54_{-0.18}^{+0.18}$ & --       & $2.90^{(e)}$    & 30.9 & 74 & 0.36 &    1.53 &   0.98 & 0.25  &    0.65  & 1.92  & 0.64 & 2.18&  30.47 \\
J2215+5135    &  2.61  & 8.40  &  $3.01_{-0.37}^{+0.33}$ & --       &$ 8.5^{(a)}$     & 31.9 & 79 & 0.30 &    1.40 &   1.03 & 0.25  &    0.80  & 1.14  & 0.61 & 2.10&  30.38 \\
J2241$-$5236  &  2.19  & 8.15  &$0.513_{-0.076}^{+0.076}$& 0.81     & $3.60^{(i)} $   & 30.0 & 77 & 0.32 &    1.44 &   1.01 & 0.25  &    0.08  & 0.91  & 0.60 & 2.00&  30.37 \\
J2302+4442    &  5.19  & 8.42  &  $1.19_{-0.23}^{+0.09}$ & --       &  $3.6^{(j)} $   & 30.6 & 68 & 0.55 &    2.03 &   0.92 & 0.30  &    0.59  & 7.27  & 0.68 & 2.18&  30.55 \\

\hline
\end{tabular}
\normalsize

Note. --- The observed quantities and model parameters of MSPs. The first column is the name of the pulsar, the second to seventh
column are the observed quantities: periods in millisecond, surface magnetic fields, distances, temperatures of two thermal components and
thermal luminosities (if available), respectively. The data of X-ray emission are taken from: (a) Gentile et al. 2013; (b) Bogdanov \& Grindlay 2009;
(c) Zavlin 2006; (d) Webb et al. 2004; (e) Ransom et al. 2011; (f) Marelli et al. 2012; (g) Pancrazi et al. 2012; (h)Wu et al. 2012;
(i) Keith et al. 2011; (j) Takahashi et al. 2012. The eighth to seventeenth column are the model parameters and derived quantities (see text for detail).
\end{center}
\end{table}
\end{landscape}

\end{document}